\documentclass{amsart}

\usepackage{amsmath,amssymb,amsthm,bm}
\usepackage[table]{xcolor}
\usepackage{dsfont}
\usepackage{hyperref}
\usepackage{physics}
\usepackage{ragged2e}
\usepackage{graphicx}
\usepackage{caption}
\usepackage{array}

\theoremstyle{plain}
\newtheorem{theorem}{Theorem}[section]

\newtheorem{lemma}[theorem]{Lemma}

\theoremstyle{definition}

\newtheorem{algorithm}[theorem]{Algorithm}

\theoremstyle{remark}
\newtheorem{remark}[theorem]{Remark}

\newcommand{\su}{\mathfrak{su}}
\newcommand{\g}{\mathfrak{g}}
\newcommand{\kalg}{\mathfrak{k}}
\newcommand{\malg}{\mathfrak{m}}
\newcommand{\halg}{\mathfrak{h}}

\newcommand{\HS}{\mathrm{HS}}
\newcommand{\Pauli}{\mathcal{P}}
\newcommand{\normop}[1]{\left\lVert #1 \right\rVert}
\newcommand{\innerHS}[2]{\left\langle #1,#2 \right\rangle_{\HS}}

\begin{document}

\title{Zassenhaus Expansion in Solving the Schr\"odinger Equation}

\author{Yali Du}
\address{Hubei University}
\email{Duyali\_123@126.com}

\author{Naihuan Jing}
\address{North Carolina State University}
\email{jing@ncsu.edu}

\author{Molena Nguyen}
\address{North Carolina State University}
\email{thnguy22@ncsu.edu}

\author{Yiling Wang}
\address{South China University of Technology}
\email{yilingw327@outlook.com}

\thanks{*Corresponding author: Naihuan Jing}


\begin{abstract}
A fundamental challenge in quantum simulation is approximating the time-evolution operator \(U(t)=e^{-i\mathcal{H}t}\) generated by a large sum of typically non-commuting Hamiltonians 
using resource-efficient circuits compatible with near-term devices. We present a refinement of fixed-depth Lie-theoretic simulation that incorporates second-order Zassenhaus commutator corrections into a Cartan/KAK decomposition template. The resulting approximation retains constant circuit depth while achieving local error \(\mathcal{O}(t^3)\) in operator norm under standard boundedness assumptions, and it substantially reduces gate counts relative to first-order product formulas when time is large and depth is constrained. The method leverages closure of Pauli commutators inside Pauli-generated Lie algebras, enabling symbolic commutator evaluation and avoiding explicit matrix exponentiation in classical preprocessing. This yields a structured pathway to compile lattice and chemistry-inspired Hamiltonians with locality constraints into fixed-depth circuits suitable for noisy intermediate-scale quantum hardware.
\end{abstract}

\maketitle

\section{Introduction}
Unitary synthesis is a central task in quantum information processing and simulation, providing a systematic procedure to compile target evolutions into elementary gates supported by a given device model~\cite{Nielsen2000, Barenco1995, Vartiainen2004, Shende2006, Motta2023, Babbush2018, Tran2023, Wang2024}. The feasibility of near-term quantum algorithms is often determined not only by asymptotic query complexity but by concrete constraints on depth, two-qubit gate counts, and classical decomposition overhead~\cite{Khatri2019, Maslov2018, Murali2023, Amy2013, Arute2019}. These concerns are particularly acute for Hamiltonian simulation, which underlies quantum chemistry~\cite{Kivlichan2018, Motta2023, Babbush2018}, condensed matter models, and many hybrid variational workflows~\cite{Peruzzo2014, Cerezo2021}.

For a time-independent Hamiltonian \(\mathcal{H}\), Schr\"odinger evolution is
\begin{equation}\label{eq:unitary-evolution}
U(t)=e^{-i\mathcal{H}t}.
\end{equation}
In many physically motivated settings, \(\mathcal{H}\) is expressed as a sum of local or quasi-local terms,
\begin{equation}\label{eq:H-decomp}
\mathcal{H}=\sum_{j=1}^m H_j,
\end{equation}
where each \(H_j\) acts on a small subset of qubits determined by an interaction graph. In general the terms do not commute, and therefore
\begin{equation}\label{eq:firstord}
e^{-i\mathcal{H}t}\neq \prod_{j=1}^m e^{-iH_j t}.
\end{equation}
Product-formula methods and their refinements provide systematic approximations of~\eqref{eq:unitary-evolution} from local exponentials~\cite{Lloyd1996, Suzuki1991, Childs2021}. Alternative simulation paradigms include Taylor-series and LCU techniques~\cite{Berry2015, Berry2022, Koch2023} and quantum signal processing~\cite{Low2017, Low2019}. While these approaches can be asymptotically efficient, depth and prefactor constraints may limit their utility on noisy intermediate-scale quantum (NISQ) devices~\cite{Preskill2018, Campbell2019}.

A complementary viewpoint uses Lie algebra structure to separate commuting and non-commuting components of the dynamics. The commutator closure of the local generators yields a dynamical Lie algebra \(\g(\mathcal{H})\subseteq \su(2^n)\), which governs reachable directions and encodes symmetries of the interaction graph~\cite{DAlessandro2007, Schirmer2001, Kokcu2023}. Cartan decompositions and related KAK factorizations are fundamental tools in quantum control and synthesis: they reduce unitary decomposition to structured conjugations around a commuting ``Cartan'' block~\cite{Khaneja2001, Nielsen2002, Shende2006, Vatan2004, Bullock2004}. This decomposition is especially useful when one seeks circuits whose depth is controlled by a chosen parameterization class, rather than by the evolution time \(t\), and it is known that The Cartan decomposition often separates the noncommuting part of $\mathcal H$ from the abelian component in an efficient way. However, the approximation of the noncommutative part is only taken in linear order, which may create error even in the microscopic magnitude. In this paper we will upgrade the algorithm by employing the Zassenhaus decomposition.

The Zassenhaus factorization expresses \(e^{A+B}\) as an ordered product whose correction factors are exponentials of nested commutators~\cite{Zassenhaus1947, Blanes2009, Casas2012}. The second-order truncation,
\begin{equation}\label{eq:z2-intro}
e^{A+B}=e^A e^B e^{-\frac12[A,B]}+\mathcal{O}(\normop{A}^3+\normop{B}^3),
\end{equation}
identifies the leading non-commutative correction missing from first-order product formulas. Specifically we incorporate these commutator corrections into a Cartan/KAK fixed-depth template so that the depth is governed by a constant-depth conjugating circuit \(K(\theta)\) and a commuting middle block, while non-commutativity is handled by analytically tractable commutators that respect locality.


The following are three technical advantages of the Zassenhaus decomposition:
\begin{itemize}
   \item It develops a Cartan (KAK-type) decomposition template designed for fixed-depth simulation. The construction specifies the chosen Cartan subalgebra, the structure of the conjugating unitaries, and the parameterized normal form that serves as the decomposition target, thereby isolating nonlocal generators while enforcing a prescribed circuit depth.
\item It incorporates multivariable Zassenhaus expansions into the parametrization to refine the local error without increasing circuit depth. The truncation order, commutator hierarchy, and resulting approximation error are stated explicitly, enabling order-by-order accuracy control.
 \item It introduces a Pauli/Lie algebra framework for symbolic commutator evaluation on the $n$-qubit Pauli basis $\{\sigma_\alpha\}_{\alpha\in\{I,X,Y,Z\}^n}$. The framework provides explicit formulas for nested commutators, identifies the associated structure constants, and tracks support growth and weight under commutation, yielding quantitative locality bounds relevant for circuit synthesis.
\end{itemize}

The paper is organized as follows.
Section~\ref{sec:cartan-template} presents the Cartan/KAK fixed-depth decomposition scheme
and formalizes the decomposition target. 
Section~\ref{sec:zassenhaus} states the multivariable Zassenhaus expansions used in the refined parameterization, including truncation order and commutator structure. 
Section~\ref{sec:prelim}
proves Pauli commutator closure properties and derives locality-aware synthesis implications for the special case of Pauli strings.
Section~\ref{sec:optimization} formulates the parameter optimization problem and details the procedure for identifying the conjugating circuit. 
Section~\ref{sec:error-complexity} establishes explicit error bounds, truncation estimates, and circuit-complexity scaling. 
Section~\ref{sec:numerics} reports numerical benchmarks, convergence behavior, and empirical scaling with system size.


\section{Fixed-depth Hamiltonian simulation using Cartan/KAK decomposition}\label{sec:cartan-template}
Let \(\g=\g(\mathcal{H})\subseteq \su(2^n)\) be the dynamical Lie algebra generated by the components of $\mathcal H$. A Cartan involution \(\Theta\) on \(\g\) yields a direct sum decomposition
\begin{equation}\label{eq:cartan-decomp}
\g=\kalg\oplus \malg,
\end{equation}
where \(\kalg\) and \(\malg\) are the \(+1\) and \(-1\) eigenspaces of \(\Theta\), respectively, satisfying
\begin{equation}\label{eq:symmetric-brackets}
[\kalg,\kalg]\subseteq \kalg,\qquad [\kalg,\malg]\subseteq \malg,\qquad [\malg,\malg]\subseteq \kalg.
\end{equation}
We choose \(\Theta\) so that \(\Theta(\mathcal{H})=-\mathcal{H}\), implying \(\mathcal{H}\in \malg\). A common involution on matrix Lie algebras is \(\Theta(X)=-X^{T}\), which satisfies \(\Theta^2=\mathrm{id}\) and induces a symmetric decomposition on appropriate subalgebras~\cite{Helgason1978, DAlessandro2007}.

A Cartan subalgebra \(\halg\subseteq \malg\) is chosen as a maximal abelian subspace, so that \(\exp(\halg)\) is a commuting torus. The KAK theorem implies that elements of the connected compact group generated by \(\g\) admit factorizations of the form
\begin{equation}\label{eq:kak}
U=K_1 A K_2,\qquad K_1,K_2\in \exp(\kalg),\ \ A\in \exp(\halg),
\end{equation}
with uniqueness up to Weyl-group symmetries~\cite{Khaneja2001, Nielsen2002, DAlessandro2007}. In synthesis, this suggests targeting a representation of the time evolution
\begin{equation}\label{eq:target-form}
e^{-i\mathcal{H}t}\approx K(\theta)^\dagger e^{-ih_0 t}K(\theta),
\end{equation}
with \(K(\theta)\in \exp(\kalg)\) of fixed depth and \(h_0\in \halg\) commuting.

The Cartan/KAK template becomes constructive once \(K(\theta)\) is restricted to a constant-depth parameterization compatible with hardware constraints~\cite{Khatri2019, Maslov2018, Murali2023}. The middle block \(e^{-ih_0 t}\) is compiled from commuting terms and can often be parallelized.

\begin{algorithm}[Fixed-depth Cartan simulation]\label{alg:fixed-depth}
Input: Hamiltonian \(\mathcal{H}\), time \(t\), and a fixed-depth parameterization family \(K(\theta)\subset\exp(\kalg)\).\\
Output: A circuit approximating \(e^{-i\mathcal{H}t}\).
\begin{enumerate}
\item Construct \(\g(\mathcal{H})\) from the Pauli support of \(\mathcal{H}\) by commutator closure~\cite{DAlessandro2007, Schirmer2001, Kokcu2023}.
\item Choose a Cartan involution \(\Theta\) and form \(\g=\kalg\oplus\malg\) with \(\mathcal{H}\in\malg\).
\item Choose \(\halg\subset\malg\) maximal abelian, and a target element \(v\in \halg\) used in the alignment objective.
\item Optimize \(\theta\) to align \(K(\theta)^\dagger v K(\theta)\) with \(\mathcal{H}\) in Hilbert--Schmidt overlap (Section~\ref{sec:optimization}).
\item Set \(K_c=K(\theta^*)\) and \(h_0=K_c \mathcal{H}K_c^\dagger\in \halg\).
\item Implement \(K_c^\dagger e^{-ih_0 t}K_c\) with commuting decomposition for the Cartan block.
\end{enumerate}
\end{algorithm}

\begin{remark}
Algorithm~\ref{alg:fixed-depth} is depth-stable in \(t\): the number of entangling layers in \(K_c\) is fixed, while the parameter values encode the target time. This aligns with NISQ constraints where depth and two-qubit errors dominate performance~\cite{Preskill2018, Campbell2019}.
\end{remark}

\section{Multivariable Zassenhaus decomposition and higher-order factorizations}\label{sec:zassenhaus}
In the usual fixed-depth Cartan simulation, the element $K(\theta)$ is approximated in linear order which still provides a quite good simulation. 
In our improved algorithm we propose to use higher-order approximation using the
Zassenhaus decomposition \cite{Zassenhaus1947}.

For non-commuting operators \(A,B\), the Zassenhaus factorization may be written as
\begin{equation}\label{eq:zassenhaus-general}
e^{A+B}=e^A e^B \prod_{k=2}^{\infty} e^{W_k(A,B)},
\end{equation}
where each \(W_k\) is a homogeneous Lie polynomial of degree \(k\) in \(A,B\) given by nested commutators~\cite{Zassenhaus1947, Blanes2009, Casas2012}. The second-order truncation is
\begin{equation}\label{eq:z2}
e^{A+B}=e^A e^B e^{-\frac{1}{2}[A,B]}+\mathcal{O}(t^3),
\end{equation}
where the error term is governed by third-order commutators such as \([A,[A,B]]\) and \([B,[A,B]]\)~\cite{Blanes2009, Casas2012, Moan2006} explicitly stated as follows.

\begin{lemma}[Second-order Zassenhaus local error]\label{lem:z2-error}
Let \(A(t)=tX\) and \(B(t)=tY\) with bounded operators \(X,Y\). Then
\begin{equation}\label{eq:z2-local}
e^{t(X+Y)}-e^{tX}e^{tY}e^{-\frac{t^2}{2}[X,Y]}=\mathcal{O}(t^3)
\end{equation}
in operator norm, where the implied constant depends on bounds for \(\normop{X}\), \(\normop{Y}\), and the relevant nested commutators.
\end{lemma}

For \(n\) non-commuting operators \(X_1,\dots,X_n\),
\begin{equation}\label{eq:multi-zassenhaus}
e^{X_1+\cdots+X_n}=e^{X_1}e^{X_2}\cdots e^{X_n}\prod_{k=2}^{\infty} e^{W_k},
\end{equation}
where \(W_k\) is a homogeneous Lie polynomial of degree \(k\) built from commutators of the \(X_i\). A recursive computation for multivariable \(W_k\) was given in~\cite{Wang2019}. 
More precisely, 
let \(k_1,\dots,k_n\in\kalg\) be generators used to define an parameterization for \(K(\theta)\). The first-order product is
\begin{equation}\label{eq:K1}
K_1(\theta)=\prod_{i=1}^n e^{i\theta_i k_i}.
\end{equation}
The second-order commutator-corrected product is
\begin{equation}\label{eq:K2}
K_2(\theta)=\left(\prod_{i=1}^n e^{i\theta_i k_i}\right)
\left(\prod_{1\le i<j\le n} e^{-\frac{1}{2}\theta_i\theta_j [k_i,k_j]}\right).
\end{equation}
Third- and fourth-order truncations follow the explicit multivariable formulas in~\cite{Wang2019}. For completeness, we record the common third-order correction written as
\begin{align}\label{eq:K3}
K_3(\theta)
&=\left(\prod_{i=1}^{n} e^{i\,\theta_i\, k_i}\right)
\left(\prod_{1 \le i < j \le n} e^{-\frac{1}{2}\,\theta_i\,\theta_j\,[k_i, k_j]}\right)\nonumber\\
&\quad\times
\prod_{1 \le i < j \le n} \exp\!\left(\frac{1}{6}\Bigl(\theta_i^2\,\theta_j\,[k_i,[k_i,k_j]] + \theta_i\,\theta_j^2\,[k_j,[k_i,k_j]]\Bigr)\right),
\end{align}
and the fourth-order term for two operators \(A,B\) in the standard form
\begin{align}\label{eq:W4}
W_4
=
-\frac{1}{24}\Bigl(
[A,[A,[A,B]]]+3[A,[B,[B,A]]]+[B,[B,[B,A]]]
\Bigr).
\end{align}
A general multivariable fourth-order truncation may be written as in~\cite{Wang2019}:
\begin{align}\label{eq:K4}
K_4(\theta)
&=\prod_{i=1}^{n} e^{i\,\theta_i\, k_i}
\prod_{1\le i<j\le n} e^{-\frac{1}{2}\,\theta_i\,\theta_j\,[k_i,k_j]}\nonumber\\
&\quad\times
\prod_{1\le i<j\le n} \exp\!\Biggl(\frac{1}{6}\Bigl(\theta_i^2\,\theta_j\,[k_i,[k_i,k_j]] + \theta_i\,\theta_j^2\,[k_j,[k_i,k_j]]\Bigr)\Biggr)\nonumber\\
&\quad\times
\prod_{1\le i<j<k<l\le n} \exp\!\Biggl(-\frac{1}{24}\,C_4(i,j,k,l)\Biggr),
\end{align}
where
\begin{align}\label{eq:C4}
C_4(i,j,k,l)
&=
[k_i,[k_j,[k_k,k_l]]]
+3[k_i,[k_l,[k_j,k_k]]]\nonumber\\
&\quad
+3[k_j,[k_k,[k_l,k_i]]]
+[k_l,[k_j,[k_k,k_i]]].
\end{align}

\begin{remark}
The practical usefulness of \(K_3\) and \(K_4\) depends on whether the additional nested commutator exponentials can be synthesized without increasing depth beyond device constraints, and on whether commutator magnitudes remain sufficiently small for truncation gains to be visible~\cite{Suzuki1990, Lloyd1996, Preskill2018, Moan2006}.
\end{remark}

\section{Pauli structure, Lie closure, and norms}\label{sec:prelim}
Let \(\Pauli_n=\{I,X,Y,Z\}^{\otimes n}\) be the set of Pauli strings on \(n\) qubits. Many Hamiltonians of interest admit sparse expansions in \(\Pauli_n\), often of bounded weight (few-body terms) and constrained by an interaction graph~\cite{Nielsen2000, Lloyd1996, Childs2021}. Specifically we can write 
\begin{equation}\label{eq:pauli-expansion}
\mathcal{H}=\sum_{\alpha} h_\alpha \sigma^\alpha,\qquad h_\alpha\in\mathbb{R},\ \sigma^\alpha\in\Pauli_n.
\end{equation}
Then 
$\g(\mathcal{H})=\mathrm{Lie}\langle i\sigma^\alpha:\ h_\alpha\neq 0\rangle \subseteq \su(2^n)$
which captures controllability and symmetry restrictions~\cite{DAlessandro2007, Schirmer2001, Kokcu2023}.

\begin{lemma}[Pauli commutator structure]\label{lem:pauli-comm}
Let \(P,Q\in \Pauli_n\). Then \(PQ=\pm QP\) and
\begin{equation}\label{eq:pauli-comm-formula}
[P,Q]=2PQ,
\end{equation}
up to an overall phase. In particular, \([P,Q]\) is proportional to a Pauli string.
\end{lemma}

Therefore
$\mathfrak g(\mathcal H)$ admits a basis consisting of Pauli strings, and commutator corrections in Zassenhaus truncations can be assembled symbolically without matrix exponentiation.
This underlies the decomposition advantage of commutator-corrected fixed-depth schemes: once a locality graph is specified, only overlapping supports contribute nonzero commutators, so the second-order factor in~\eqref{eq:z2} can be restricted to local neighborhoods~\cite{Childs2021, Cirstoiu2020, Wiersema2020}.

We use the operator 2-norm \(\normop{\cdot}_2\) to quantify simulation error and to state stability bounds for exponentials and splittings~\cite{Higham2002}. For orbit optimization and alignment objectives we use the Hilbert--Schmidt inner product
\(
\innerHS{A}{B}=\Tr(A^\dagger B),
\)
which is standard in Lie-theoretic control~\cite{DAlessandro2007, Schirmer2001}.

If \(\mathcal{H}\) is a sum of local terms supported on edges or small hyperedges of an interaction graph, then commutators \([H_i,H_j]\) vanish unless the supports overlap. This sparsity reduces both classical preprocessing and the number of distinct commutator exponentials that must be synthesized in a second-order truncation~\cite{Childs2021, Haah2021}. In the Cartan/KAK template~\eqref{eq:target-form}, commutator corrections can be absorbed into the fixed-depth conjugating layers or into the Cartan block depending on the chosen parameterization and decomposition, with synthesis guided by two-qubit canonical decompositions when working at small locality~\cite{Vatan2004, Bullock2004, Shende2006}.

\section{Optimization of the conjugating transformation}\label{sec:optimization}
We now present a smooth optimization procedure that identifies a conjugating circuit \(K(\theta)\) bringing \(\mathcal{H}\) into the Cartan sector.

Let \(\halg\subset\malg\) be a fixed maximal abelian subspace spanned by \(\{h_i\}\). We choose a generic target element (called {\it Cartan target})
\begin{equation}\label{eq:v-target}
v=\sum_i \gamma^i h_i,
\end{equation}
where the real coefficients \(\gamma^i\) are taken irrational (often transcendental, such as \(\pi\)) to reduce degeneracies in the orbit objective~\cite{SchulteHerbruggen2005}. This choice makes the alignment landscape less symmetric and improves optimizer identifiability in practice.

As objective, we seek \(\theta\in\mathbb{R}^d\) such that \(K(\theta)^\dagger vK(\theta)\) aligns with \(\mathcal{H}\) in Hilbert--Schmidt overlap:
\begin{equation}\label{eq:cost}
f(\theta)=\Tr\!\left(K(\theta)^\dagger v K(\theta)\,\mathcal{H}\right).
\end{equation}
This objective selects a point on the conjugacy orbit of \(v\) that best matches \(\mathcal{H}\) under the \(\HS\) pairing, a standard construction in Lie-theoretic control and synthesis~\cite{DAlessandro2007, Schirmer2001, Dirr2008}. The parameterization \(K(\theta)\) is built from Zassenhaus-based products \(K_p(\theta)\) (Section~\ref{sec:zassenhaus}), with \(p=2\) the default choice in our numerical studies.

We use BFGS (Broyden–Fletcher–Goldfarb–Shanno algorithm) (or its limited-memory variant) to optimize~\eqref{eq:cost}~\cite{Nocedal2006}. The quasi-Newton optimization uses gradient information and an approximate inverse Hessian updated iteratively. In circuit implementations, gradients may be estimated by the parameter-shift rule or finite-difference surrogates; in classical benchmarking, gradients can be computed directly from matrix expressions. The output is an optimizer \(\theta^*\) yielding \(K_c:=K(\theta^*)\).

Once \(K_c\) is identified, define
\begin{equation}\label{eq:h0-def}
h_0=K_c \mathcal{H}K_c^\dagger\in \halg.
\end{equation}
Then the compiled simulation is
\begin{equation}\label{eq:compiled-U}
U(t)=e^{-i\mathcal{H}t}\approx K_c^\dagger e^{-ih_0 t}K_c,
\end{equation}
where \(e^{-ih_0 t}\) is synthesized from commuting exponentials. The depth is determined by the fixed parameterization for \(K_c\).

\section{Error, fidelity, and complexity considerations}\label{sec:error-complexity}
The second-order Zassenhaus truncation improves local error order compared to first-order product formulas by including the leading commutator correction~\cite{Blanes2009, Casas2012, Moan2006}. For \(H=A+B\), one may compare Lie--Trotter splitting vs. commutator-corrected splitting:
\begin{align}\label{eq:ltrotter}
e^{-iHt}&\approx \left(e^{-iAt/m}e^{-iBt/m}\right)^m,\qquad \text{error } \mathcal{O}(t^2/m),\\
\label{eq:z2-splitting}
e^{-iHt}&\approx \left(e^{-iAt/m}e^{-iBt/m}e^{i t^2[A,B]/(2m^2)}\right)^m,\qquad \text{error } \mathcal{O}(t^3/m^2),
\end{align}
consistent with Lemma~\ref{lem:z2-error} and standard splitting analyses~\cite{Berry2015, Childs2021, Haah2021}. In regimes where depth prohibits taking \(m\) large, a fixed-depth scheme benefits from incorporating commutator structure directly into the compiled blocks.

Commutator-aware factorizations tend to better preserve spectral and geometric features that can be distorted by low-order splitting, particularly over long times or under strong interactions~\cite{Vidal2004, Sahinoglu2020, Higham2002}. In Lie-theoretic decomposition, maintaining the group geometry of the evolution is important for stable synthesis, and commutator corrections align naturally with the KAK template because they remain within the same Lie closure~\cite{Helgason1978, Nielsen2002}.

Now let's study gate complexity and depth comparison.
Let \(H=\sum_{\ell=1}^{m} h_\ell\) be a local Hamiltonian supported on an interaction graph \(G=(V,E)\) of maximum degree \(\Delta\), with chromatic index \(\chi'(G)\le \Delta+1\). In a depth-optimized first-order product formula, a single step may be scheduled using an edge-coloring so that two-qubit layers correspond to colors. With step size \(\Delta t\) and \(r=t/\Delta t\), the depth scales as \(\chi'(G)\,r\) and the two-qubit gate count scales as \(|E|\,r\)~\cite{Suzuki1990, Lloyd1996, Childs2021}.

In the fixed-depth Cartan/Zassenhaus approach, the circuit has the form \(K_c^\dagger e^{-ih_0 t}K_c\). Let \(D_K\) denote the two-qubit depth of \(K_c\) and \(D_A\) denote the two-qubit depth of the commuting Cartan block (including any local commutator exponentials required by the chosen Zassenhaus truncation). Then
\begin{equation}\label{eq:depth-z2}
\text{depth}_{\mathrm{Z2}}=2D_K+D_A,
\end{equation}
which is independent of the number of time segments \(r\). If the parameterization respects hardware locality, then \(D_K\) and \(D_A\) are constant on fixed-degree architectures, yielding a constant-depth simulation in \(t\)~\cite{Preskill2018, Cirstoiu2020, Wiersema2020}.

As for the optimization cost, instead of optimizing over \(\su(2^n)\) (dimension \(4^n-1\)), we restrict \(K(\theta)\) to a constant-depth parameterization with parameter count \(p=O(n)\). Using L-BFGS with fixed history length, per-iteration classical work scales as \(O(p)\) and memory as \(O(p)\)~\cite{Nocedal2006}. Symbolic commutator assembly is efficient because nonzero commutators occur only for overlapping local supports and remain in the Pauli span. 

\section{Numerical benchmarks and scaling}\label{sec:numerics}

\begin{figure*}[!htbp]
    \centering
    
    \includegraphics[height=0.47\textheight, width=1.07\textwidth]{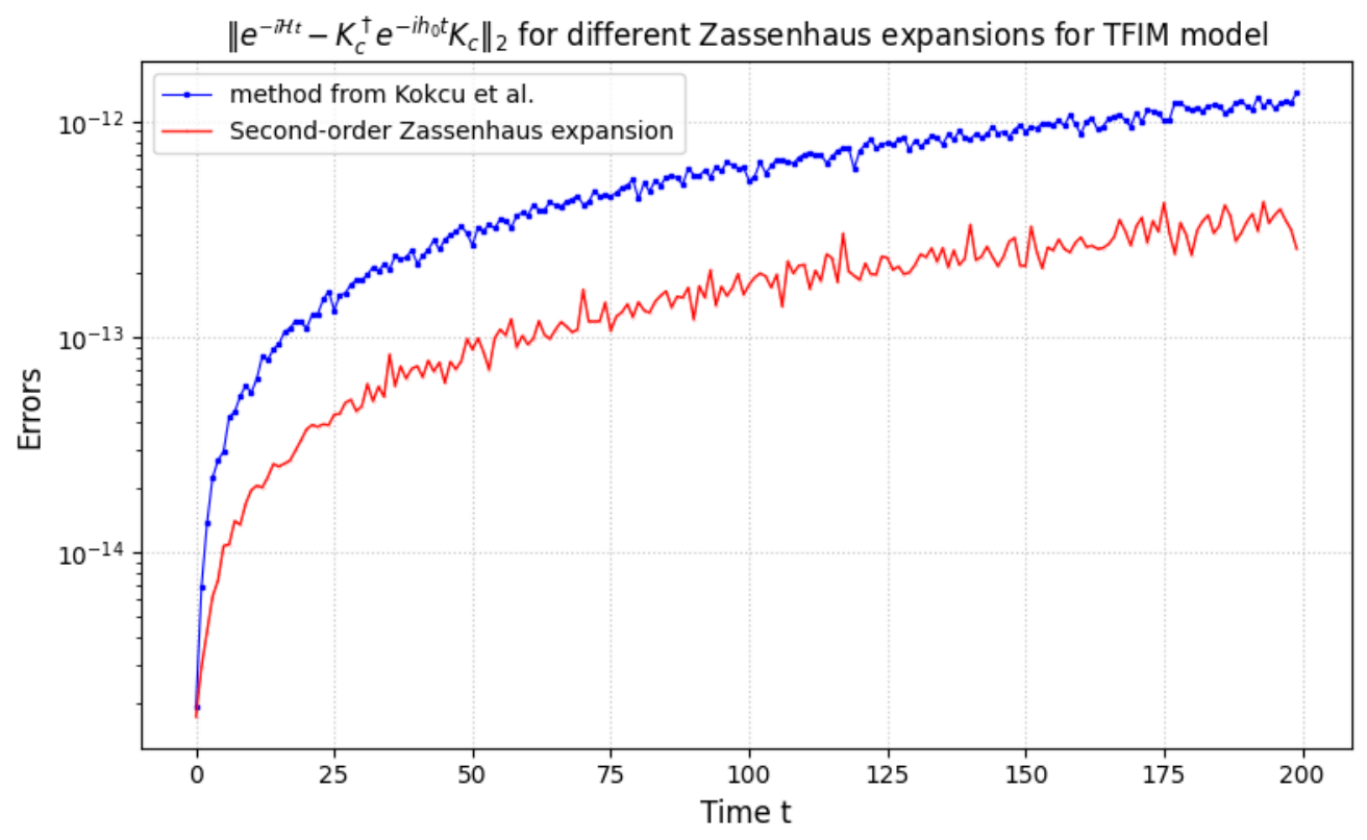}
    \\[1em]    Figure (a)
    
    \vspace{2em}
    
    \includegraphics[height=0.47\textheight,width=1.07\textwidth]{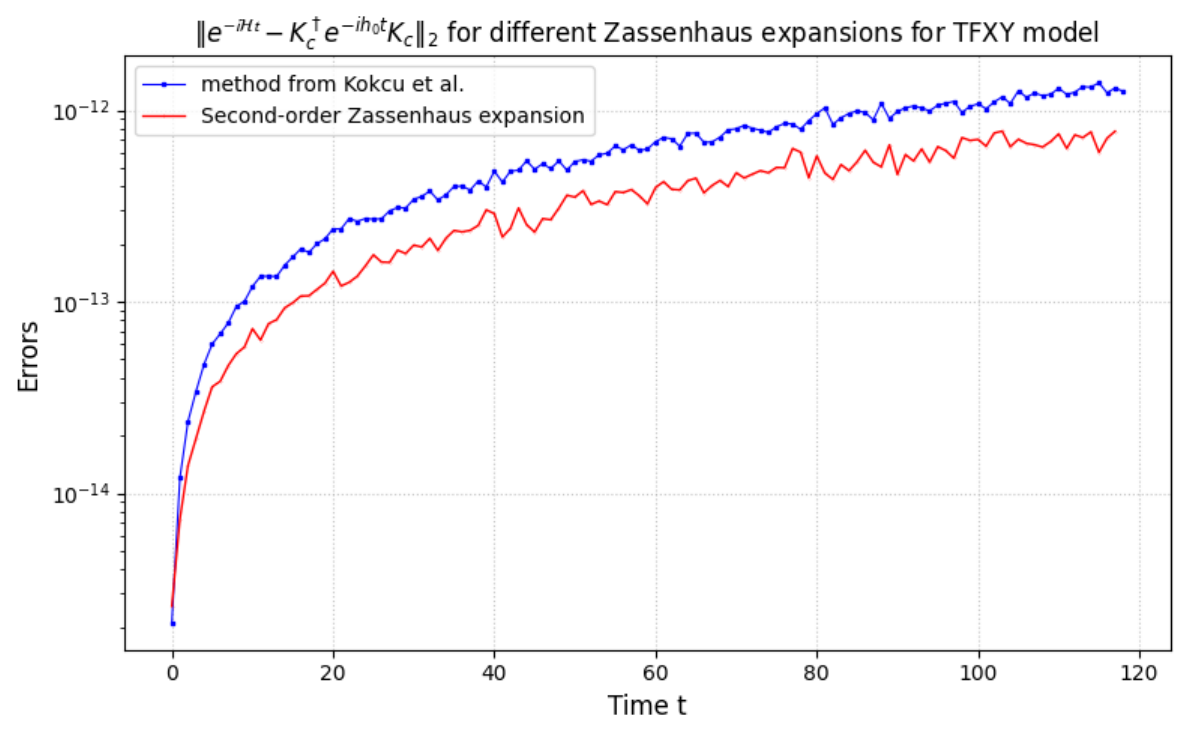}
    \\[1em] Figure (b)
    
    \caption{Two-norm error $\|e^{-i\mathcal{H}t} - K_c^\dagger e^{-ith_0t}Kc\|_2$
    \\for approximate evolutions using two different methods 
    \\for Figure 
    (a) the \texttt{TFIM} model 
      and Figure
    (b) the \texttt{TFXY} model.}
    
    \label{fig:results}
\end{figure*}

Let's compute the operator 2-norm error
\begin{equation}\label{eq:error-metric}
\left\| e^{-i \mathcal{H} t} - K_c^\dagger e^{-i h_0 t} K_c \right\|_2,
\end{equation}
over a time window \(0\le t\le 200\) for representative spin models, including the Transverse-field Ising model (\texttt{TFIM}) and the Transverse Field XY (\texttt{TFXY}) models. The baseline fixed-depth comparator is the Cartan-based method of~\cite{kokcu2023cartan}. The refined method differs only by the inclusion of the second-order commutator corrections in the construction of \(K(\theta)\) and the associated compiled blocks.

In Fig.~\ref{fig:results}(a)--(b), the second-order Zassenhaus refinement yields uniformly smaller errors over the tested window for the shown models. In these regimes, higher-order truncations beyond second order yield comparatively small additional improvements relative to the increased synthesis burden, consistent with standard splitting tradeoffs for nested commutators~\cite{Moan2006, Casas2012, Blanes2009, Childs2021, Haah2021}.

\begin{table}[ht]
\centering
\caption{Comparison of the second-order Zassenhaus refinement with the baseline fixed-depth method of K\"okc\"u \textit{et al.}~\cite{kokcu2023cartan}.}
\label{tab:comparison}
\begin{tabular}{l>{\centering\arraybackslash}p{3cm}
                  >{\centering\arraybackslash}p{3cm}}
\hline
Model & Baseline & 2nd Zassenhaus \\
\hline
\texttt{TFIM}           & \(7.63\times 10^{-14}\)  & \(\mathbf{2.86\times 10^{-14}}\) \\
\texttt{TFXY}           & \(1.25\times 10^{-13}\)  & \(\mathbf{7.66\times 10^{-14}}\) \\
\texttt{Heisenberg}     & \(1.008\times 10^{-13}\) & \(\mathbf{9.93\times 10^{-14}}\) \\
\texttt{XY}             & \(3.01\times 10^{-14}\)  & \(3.01\times 10^{-14}\) \\
\texttt{Kitaev\_even}   & \(1.63\times 10^{-14}\)  & \(1.63\times 10^{-14}\) \\
\texttt{Kitaev\_odd}    & \(1.63\times 10^{-14}\)  & \(1.63\times 10^{-14}\) \\
\hline
\end{tabular}
\end{table}

Table~\ref{tab:comparison} records representative two-norm errors. The second-order Zassenhaus refinement reduces error relative to the baseline for \texttt{TFIM}, \texttt{TFXY}, and \texttt{Heisenberg}, while \texttt{XY} and the listed Kitaev instances exhibit no change at the numerical precision shown. Such plateaus are consistent with regimes where the dominant error is already at a numerical floor or where commutator corrections vanish or are negligible on the tested instances.

Let \(H_{\mathrm{hw}}=(V_{\mathrm{hw}},E_{\mathrm{hw}})\) be the hardware connectivity graph with chromatic index \(\chi'(H_{\mathrm{hw}})\le \Delta_{\mathrm{hw}}+1\). If \(R(K)\) is the maximum graph distance spanned by any entangling interaction in \(K(\theta)\), then a simple scheduling bound gives
\begin{equation}\label{eq:DK-bound}
D_K \le \chi'(H_{\mathrm{hw}})\,R(K).
\end{equation}
Thus, on fixed-degree nearest-neighbor architectures with constant-range entanglers, \(D_K=O(1)\), and the total depth in~\eqref{eq:depth-z2} is constant. On architectures requiring routing (e.g.\ long-range interactions on a 1D chain), effective range growth can induce \(D_K\) scaling consistent with SWAP-network lower bounds.

\section{Conclusion}
We presented an improved Cartan/KAK fixed-depth Hamiltonian simulation via Zassenhauss decomposition. By incorporating second-order commutator corrections into a constant-depth decomposition template, the method enhances local accuracy while preserving a synthesis structure compatible with NISQ constraints. We have applied the new algorithm to models involving Pauli commutator closure to enable symbolic preprocessing and locality-aware commutator placement which also reduces decomposition overhead. The resulting framework provides a Lie-theoretic pathway to depth-limited simulation for structured Hamiltonians, with empirical improvements demonstrated on benchmark spin models.

\section*{Acknowledgement}
We thank Lex Kemper for helpful discussion and the Simons Foundation for partial support.


\end{document}